\documentclass[aps,prd,twocolumn,showpacs,preprintnumbers,nofootinbib,amsmath,amssymb,floatfix]{revtex4}

\usepackage{graphicx}
\usepackage{dcolumn}
\usepackage{bm}
\usepackage{amsmath}

\newcommand{\beq}{\begin{eqnarray}}
\newcommand{\eeq}{\end{eqnarray}}
\newcommand{\ie}{{\it i.e.\ }}
\newcommand{\eg}{{\it e.g.\ }}

\begin{document}

\title{The three-dimensional, three state Potts model in a negative external field}

\author{Claudio Bonati$^1$ and Massimo D'Elia$^2$}

\affiliation{$^1$Dipartimento di Fisica, Universit\`a di Pisa and INFN, Largo Pontecorvo 3, I-56127 Pisa, Italy\\
$^2$Dipartimento di Fisica, Universit\`a di Genova and INFN, Via Dodecaneso 33, 16146 Genova, Italy}
\date{\today}

\begin{abstract}
We investigate the critical behaviour of the three-dimensional, three state Potts
model in presence of a negative external field $h$, \ie disfavouring one
of the three states. A genuine phase transition is present for all values of $|h|$, 
corresponding to the spontaneous breaking of a residual
$Z_2$ symmetry. The transition is first/second order respectively for 
small/large values of $|h|$, with a tricritical field $h_{\rm tric}$ separating the 
two regimes.
We provide, using different and consistent approaches, a precise determination 
of $h_{\rm tric}$; we also compare with previous studies and 
discuss the relevance of our investigation to analogous studies of the QCD phase diagram
in presence of an imaginary chemical potential.  
\end{abstract}

\pacs{
64.60.Kw (Multicritical points),
75.10.Hk (Classical spin models), 
11.15.Ha (Lattice gauge theory).
}
\maketitle

\section{Introduction}

Potts models~\cite{potts} have been often considered in literature to reproduce the critical
properties of more complex physical 
systems~\cite{wu,svya,gavai,fuku,alves,schmidt,Janke,Karsch,caselle1,pottsim1,pottsim2,caselle2,papa,babe,babedu}.
In the present work we are interested in the 3-state Potts model defined on a three dimensional cubic lattice.
The generic $q$-state Potts model is defined by the following partition function
\beq
Z(\beta, H) = \sum_{\{ \sigma_i \}} {\rm e}^{-\beta (E - H M)} \, ,
\label{Z}
\eeq
where the spin variable $\sigma_i$ lives on lattice site $i$ and can 
take $q$ different possible values,
\eg $\sigma_i \in \{ 0,1, \dots ,(q-1) \}$, while $\beta = 1/(k_B T)$.
$E$ and $M$ denote respectively the energy and magnetization with respect 
to a chosen reference spin value $\bar\sigma$ (\eg $\bar\sigma = 0$):
\beq
E &=&  - J \sum_{\langle i,j \rangle} \delta_{\sigma_i, \sigma_j} \label{energy} \\
M &=& \sum_i \delta_{\sigma_i, \bar\sigma}
\label{EM}
\eeq
where $J$ is the coupling constant, $H$ is an external applied field
and $\langle i,j \rangle$ in the sum denotes all pairs of nearest neighbor
lattice sites.
In the following,
as usual, we shall set $J = 1$ and make use of the normalized magnetic field 
$h \equiv \beta H$.

For $h = 0$ the system has an exact symmetry, corresponding 
to all possible global permutations of the $q$ spin values,
\ie the symmetry group is the group of permutations 
$S_q$. Such symmetry gets spontaneously broken below
a given critical temperature,
where the spin variables align themselves along a given direction.
The corresponding phase transition is first order, in three dimensions,
for $q \geq 3$, and second order for $q = 2$ (the system coincides
with the Ising model in this case). For $q = 3$ the critical temperature 
is given by $\beta_c(h=0) = 0.550565(10)$ \cite{Janke}.

The critical properties of 
3D Potts models with $q = 2$ or 3 at $h = 0$, have often been associated with those
met at the finite $T$ phase transition of QCD (with 2 or 3 colors) in 
the pure gauge limit, via the well known Svetitsky-Yaffe conjecture~\cite{svya}.
The symmetry group which is spontaneously broken in the high T, deconfined phase of 
$SU(N)$ pure gauge theories is that associated with center symmetry, $Z_N$,
corresponding to local gauge transformations which are periodic 
in the Euclidean time direction only up to a global group element belonging
to the center of the gauge group.    
The symmetry group coincides with the 
permutation group for $N = 2$, while for $N = 3$ one has to add charge conjugation to 
center transformations to recover the full permutation group $S_3$.
The corresponding order parameter in pure gauge theories, playing the role
of magnetization and signalling the spontaneous breaking of center symmetry 
in the deconfined phase, is the Polyakov loop, \ie  a closed parallel transport in the Euclidean
time direction.

For $h \neq 0$ the symmetry $S_q$ is explicitly broken to a residual
$S_{q-1}$, corresponding to permutations among spin values other than
the chosen direction $\bar\sigma$. The case $h > 0$, in which alignment of 
spin variables along $\bar\sigma$ is favoured, has been extensively studied
in the literature: one can still distinguish two phases in which the system
is more (low T) or less (high T) aligned along $\bar\sigma$, however 
in both phases the residual $S_{q-1}$ symmetry stays unbroken, so that 
no real phase transition is expected a priori. However, for $q \geq 3$,
the first order transition present at $h = 0$ persists also for 
non zero positive values of $h$, till a critical endpoint is met, after
which the transition disappears. Such critical endpoint is expected to be
in the Ising 3D universality class and for $q = 3$ it has been located at
$(\beta_c, h_c) =(0.54938(2), 0.000775(10))$~\cite{Karsch, pottsim2}.

In the analogy with finite $T$ pure gauge theories, the case $h > 0$ corresponds to 
adding dynamical fermions of mass $m$ and in the fundamental representation of the gauge group
(the limit $h \to 0$ corresponding to $m \to \infty$): that induces an effective 
coupling to the Polyakov loop, which breaks center symmetry and aligns the Polyakov
loop along the positive real direction, while the residual charge conjugation symmetry
stays unbroken for all values of $T$.

The case $h < 0$ is quite different. Indeed in this case the reference state
$\bar\sigma$ is disfavoured, so that low T ordering happens along one of the remaining
$q-1$ states: there is therefore an order/disorder transition associated with the 
spontaneous breaking of the unbroken $S_{q-1}$ symmetry group. Actually, in the limit
of large $|h|$, the system becomes completely equivalent to a $(q - 1)$-state
Potts model at zero magnetic field, since the disfavoured state disappears from the 
statistical ensemble. Therefore a true phase transition is expected for every value
of $|h|$, coinciding with the transition of $q$- or $(q-1)$-state Potts model in the limit
of zero or infinite field respectively.

In the present paper we shall discuss the case $q = 3$ in three dimensions,
which is particularly interesting (as well as the case $q = 5$ in two dimensions), 
since in this case the transition
at $h = 0$ is first order, while the transition at $h = -\infty$ is second order in the 3D Ising
universality class. Hence the expectation is that the first order continues
for small values of $|h|$, until a tricritical point $h_{\rm tric}$ is met, governed by mean field
indexes, after which the transition becomes second order in the 3D Ising universality class.
An accurate verification of this scenario and the precise location of the tricritical point
is the aim of our study.
Notice that, for small $|h|$, we expect an interesting example of system which may
be naively believed in the Ising 3D universality class because of symmetry reasons
(the relevant symmetry being $Z_2$), but has instead a first order transition because
of the interplay with different dynamical degrees of freedom, corresponding to
the disfavoured state $\bar\sigma$ in this case.

Going back to the correspondence with the critical properties of $SU(3)$ lattice gauge 
theories, switching the sign of $h$ is like turning the boundary
conditions of dynamical fermions in the temporal direction from anti-periodic
to periodic: fermions are not thermal any more and the Euclidean temporal direction 
can then be viewed as a compactified spatial direction; the effective coupling 
to the Polyakov line changes sign, so that the Polyakov line tends to align along 
one of the complex center elements below a given compactification radius, thus
breaking spontaneously the residual charge symmetry (see \eg Refs.~\cite{degrand0,degrand,lucini,lucini2} 
for early lattice studies of the associated transition, which has been studied in the context of orientfold planar equivalence 
\cite{asv, uy2006}). Alternatively, one can interpret
the system as the usual thermal theory in presence of a purely imaginary quark number
chemical potential such that ${\rm Im}(\mu)/T = \pi$: in that case the $Z_2$ breaking 
transition is interpreted as the endpoint of the high $T$ Roberge-Weiss (RW) transitions
which are met in the $T$-${\rm Im}(\mu)$ plane~\cite{rw}. 

The importance of this $Z_2$ transition and of its order for the general features of the QCD phase diagram has been
discussed extensively in recent literature~\cite{sqgp,Kouno:2009bm,rwep,rwep2,aarts,ssky}.
In particular, its order has been investigated by lattice simulations in QCD with
two degenerate flavors in Ref.~\cite{rwep}, and more recently also for the 
three flavor theory~\cite{rwep2}: in both cases one finds a non-trivial phase structure,
with the transition being first order both for small and high quark masses, and second order
in the middle. Such phase structure can be mapped to that of the Potts model with a negative
magnetic field, which is the subject of our study, on the large mass side; on the other hand,
on the small quark mass side, chiral degrees of freedom come into play, requiring a different 
effective model description. In the context of the investigation of the QCD phase diagram, 
it is of course particularly important to give precise estimates of the tricritical values of the quark 
mass, separating the second order from the first order regions.

The study of the 3D three-state Potts model in a negative magnetic field can be placed in the more
general context of studies of the same model in complex magnetic fields~\cite{pottsim1,pottsim2,rwep2},
aimed at mimicking the dynamics of QCD in presence of a quark chemical potential, which have 
also considered the properties of the tricritical point~\cite{rwep2}. Our purpose is that of performing
a detailed study of the critical behavior of the system as a function of $h$, with the specific 
aim of determining the location of the tricritical field $h_{\rm tric}$.
We will make use of different and consistent approaches in order to do that: the strategy 
developed for this model and the corresponding results can then be taken as a guideline 
for the analogous determination of the tricritical masses for the endpoint of the 
Roberge-Weiss transition in QCD \cite{bcds}.

The paper is organized as follows: in Sec.~\ref{setup} we present and discuss the different
strategies used to investigate the critical properties of the system; in Sec.~\ref{results}
we present our numerical results and finally, in Sec~\ref{concl}, we give our conclusions.

\section{Observables and numerical analysis setup}\label{setup}

\begin{table}[bt!]
\begin{tabular}{|c|c|c|c|c|c|}
\hline & $\nu$ & $\gamma$ & $\alpha$ & $\gamma/\nu$ & $\alpha/\nu$\\
\hline $3D$ Ising & 0.6301(4) & $1.2372(5)$ & 0.110(1) & $\sim 1.963$ & $\sim 0.175$ \\
\hline Tricritical & 1/2 & 1 & 1/2 & 2 & 1\\
\hline $1^{st}$ Order & 1/3 & 1 & 1 & 3 & 3\\
\hline
\end{tabular}
\caption{Critical exponents (see \eg \cite{landau, pelissvic}).}\label{CRITEXP}
\end{table}

Natural observables for the Potts model are the energy $E$, which is defined in Eq.~(\ref{energy}), 
and the magnetization. As for the latter, we replace it by a new quantity $P$ which, in the analogy with 
QCD, plays the role of the average Polyakov line. In order to define $P$, let us associate with each spin 
variable a complex number on the unit circle as follows
\beq
s_i = \exp\left( \frac{i 2 \pi \sigma_i}{3} \right) \, ;
\eeq
then we define
\beq
P = \frac{1}{V} \sum_i s_i \, ,
\eeq
where $V = L^3$ is the lattice volume. Assuming that the state coupled to the magnetic field is 
$\bar\sigma = 0$, the residual $Z_2$ symmetry of the model corresponds to an exchange of the 
states $1$ and $2$, \ie to complex conjugation for the complex spin variables $s_i$ and for $P$.
Therefore, while $E$ and ${\rm Re} (P)$ are even under the residual $Z_2$ symmetry,
${\rm Im} (P)$ is odd and plays the role of the order parameter for the realization 
of this symmetry.

The purpose of our investigation is that of determining the location of the phase transition, 
its order and universality class, as a function of the magnetic field $h$, which is taken to 
be negative. In order to do that, we shall consider at first the susceptibility of the order
parameter
\beq
\chi \equiv L^3\ (\langle {\rm Im}(P)^2 \rangle - \langle |{\rm Im}(P)| \rangle^2) \label{susc}
\eeq 
and the specific heat of the system
\beq
C \equiv L^3\ (\langle E^2 \rangle - \langle E \rangle^2) \, .
\eeq
The scaling of the two quantities around the phase transition, as a function of the size
of the system, is fixed by the respective critical indexes
\beq
\chi = L^{\gamma/\nu}\ \phi_1 (t L^{1/\nu}) \, \label{fss1}
\eeq
and
\beq
C = C_0 + L^{\alpha/\nu}\ \phi_2 (t L^{1/\nu}) \, ,  \label{fss2}
\eeq
where $C_0$ is a regular contribution and $t \equiv (T - T_c)/T_c$ is the reduced temperature. 
In the following we will be interested in particular in the scaling of the height of the peaks, 
which is regulated by $\gamma/\nu$ and $\alpha/\nu$ respectively, and in the scaling of the width 
of the peaks, which is regulated by $1/\nu$ in both cases.
The critical indexes which are relevant to the different possibilities which may take place (\ie 
first order, second order in the 3D Ising universality class and mean field tricritical) are listed 
in Table~\ref{CRITEXP}. 

Another interesting quantity is the modulus of $P$, however it takes contribution both from
the order parameter and from the spin state coupled to the magnetic field, which is 
$Z_2$ even, hence its susceptibility is expected to be the mixing of different contributions 
scaling like $\chi$ or $C$ respectively, apart from the limit $h \to \infty$, 
in which case the contribution of the state coupled to $h$ is completely 
suppressed and the system can be mapped exactly to a 3D Ising model. 
We shall not consider such quantity in the following.

\begin{figure}[h!]
\includegraphics*[width=0.46\textwidth]{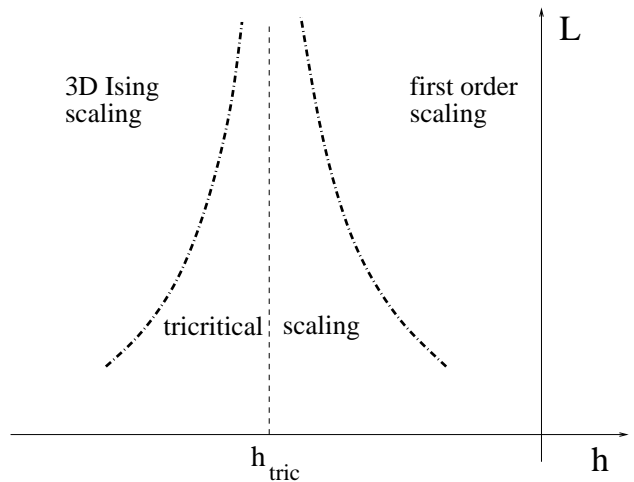}
\caption{On finite volumes tricritical scaling is expected to dominate
a finite range of $h$ values around $h_{\rm tric}$, which shrinks to zero as $L \to \infty$.}
\label{triscal}
\end{figure}

It is interesting to notice that, while first order scaling and 3D Ising scaling are expected 
to take place for a continuous range of values of $h$, tricritical scaling is in principle expected 
only for a specific value $h_{\rm tric}$, which we want to determine, at the boundary between the 
first order and the second order region. 
However the correct expectation is to have tricritical scaling regulating a neighborhood
of $h_{\rm tric}$, with the size of the neighborhood going to zero as $L \to \infty$. Putting the question
the other way around, we expect to need increasingly large volumes to discriminate between 
first order and 3D Ising second order as we approach the tricritical field $h_{\rm tric}$ from either side, 
since a fictitious tricritical scaling will mask the true thermodynamical limit for not large enough
volumes (see Fig.~\ref{triscal} for a graphical representation of that). 

This reasoning can be made more quantitative by use of the so-called crossover exponents: in the thermodynamical
limit, the true critical behaviour of the system can be seen only for $|t|\lesssim p^{1/\phi}$, where $t$ is the 
reduced temperature, $p$ is the parameter that controls the change of critical behaviour and $\phi$ is the 
crossover exponent (see \eg \cite{Cardy, BinderDeutsch, pelissvic}), which is by definition $\phi=y_p/y_t$ ($y_t$ and $y_p$ 
are the renormalization group eigenvalues of the relevant variables $t$ and $p$). In our case $p\propto h-h_{\rm tric}$ 
and $\phi=1/2$~\cite{LawSarb}. On a finite lattice of typical size $L$,
$|t|$ can be traded for $L^{-1/\nu}$ and the previous condition becomes 
$L\gtrsim |h-h_{\rm tric}|^{-\nu/\phi}$; in particular, according to the known tricritical indexes in Table~\ref{CRITEXP}, 
one expects tricritical behaviour to dominate up to a critical size 
\beq
L_c \simeq A\ |h - h_{\rm tric}|^{-1}
\label{critsize}
\eeq
where $A$ is some unknown 
constant which may be different on the first order and on the second order side; a numerical check of this behaviour will
be reported in Sec.~\ref{resultsA} (see in particular Fig.~\ref{triscal_data}). 
That implies that a correct and precise determination of $h_{\rm tric}$ may be quite 
difficult if one looks at the finite size scaling of susceptibilities or other quantities alone.

\begin{figure}[t!]
\includegraphics*[width=0.47\textwidth]{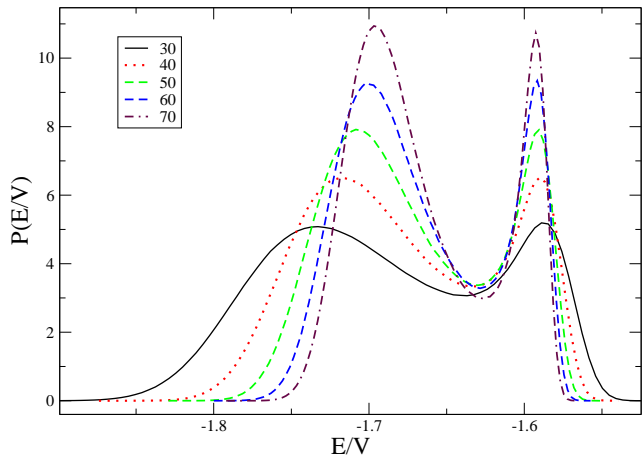}
\caption{Probability distribution of the energy density for different lattice sizes $L$ and $h = -0.0025$, where the transition is first order.}
\label{histo0.0025}
\end{figure}

\begin{figure}[t!]
\includegraphics*[width=0.47\textwidth]{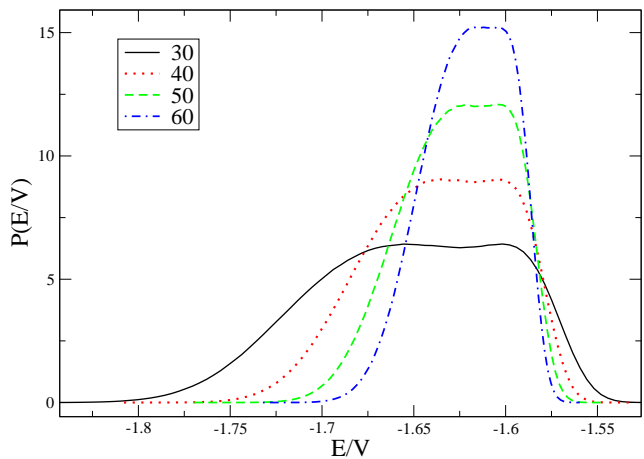}
\caption{Probability distribution of the energy density for different lattice sizes and $h = -0.01$, 
where the transition is second order.}
\label{histo0.01}
\end{figure}

As an alternative and easiest way to determine $h_{\rm tric}$, we shall determine quantities which 
give a measure of the strength of the first order transition, such as the latent heat or the 
gap of the order parameter at the transition, and study the variation of such quantities as a 
function of $h$, in order to extrapolate the point $h_{\rm tric}$ where they vanish, \ie where the first 
order disappears, without the need of making simulations very close to $h_{\rm tric}$.

We shall consider in particular the Binder-Challa-Landau cumulant~\cite{Challa} of the energy,
which is defined as \(B_4=1-\langle E^4\rangle/(3\langle E^2\rangle ^2)\). It 
can be shown (see \eg~\cite{LeeKosterlitz}) that near a transition \(B_4\) develops minima 
whose depth scales as
\begin{eqnarray}
B_4|_{min} &=&
\frac{2}{3}-\frac{1}{12}\left(\frac{E_+}{E_-}-\frac{E_-}{E_+}\right)^2+O(L^{-3}) \nonumber \\
&=&
\frac{2}{3}-\frac{1}{3}\left(\frac{\Delta_E}{\epsilon}\right)^2+O(\Delta_E^3)+O(L^{-3})
\label{Bmin}
\end{eqnarray}
where \(E_{\pm}=\lim_{\beta\to\beta_c^{\pm}}\langle E\rangle\), \(\Delta_E=E_+-E_-\) and 
\(\epsilon=\frac{1}{2}(E_++E_-)\). In particular the thermodynamical limit of  \(B|_{min}\) is less 
than \(2/3\) if and only if a latent heat is present; to simplify the notation in the following we will use 
the shorthand $B=\frac{2}{3}-B_4|_{min}$.

A different, but analogous quantity is the gap of the order parameter, $\Delta$, which can be extracted by 
looking at the scaling of the maximum of its susceptibility, $\chi$, and using the relation, valid in the 
large volume limit for a first order transition,
\beq
\chi_{\rm max} \sim A +  \frac{L^3}{4} \Delta^2 \, . \label{suscmax}
\eeq

Both $\Delta_E$ and $\Delta$ are expected to vanish as we approach the tricritical field
$h_{\rm tric}$ from the first order side. In particular, the leading order expected behaviour is the following
(see \cite{LawSarb} or \cite{Sheehy} for a brief summary)
\beq
\Delta_E \propto \sqrt{h - h_{\rm tric}} \label{deltae_beh}
\eeq
and
\beq
\Delta \propto \sqrt{|(h - h_{\rm tric})\log(h-h_{\rm tric})|} \label{delta_beh}
\eeq

Another useful quantity is the fourth-order cumulant of the order parameter. This is usually  defined by 
$\langle M^4\rangle/\langle M^2\rangle^2$ (\cite{Binder1, Binder2}), where $M$ is the order parameter, and is
typically used in the study of second order transitions. Since in this work we will analyze mainly region of the parameter 
space in which first order transitions are present, the connected form
\beq
U_4=\frac{\langle (\delta M)^4 \rangle}{\langle (\delta M)^2\rangle ^2} \qquad 
\delta M=| {\rm Im}P|- \langle|{\rm Im}P|\rangle  \label{z2_binder}
\eeq
appears to be best suited to disentangle the fluctuations inside a thermodynamical phase from the tunneling between the 
two sectors with different $Z_2$ magnetization.
Another reason to prefer the connected form is that at the critical point it develops a minumum, making thus possible to obtain the value of the cumulant 
at the transition, without introducing cross-correlations with other observables, or between different lattice size data. As a last point, we note that the relative    
error of the cumulant value at transition turned out to be smaller by a factor ~2 for the connected cumulant than for the usual one. 

For a first order transition it is simple to show that in the thermodynamic limit $U_4\to 1$, by using a double gaussian 
approximation for the distribution of the order parameter.
For a second order transition it can be shown that the value of $U_4$ at the transition is a renormalization
group invariant (\cite{Binder1, Binder2}), so that the intersection point of $U_4$ calculated on two 
lattices of different size can be used as an estimator of the transition point. It can also be shown 
that the slope of $U_4$ at the transition point $U_4^*$ satisfies the relation
\beq
\left. \frac{\partial U_4(bL)}{\partial U_4(L)}\right|_{U_4^*}=b^{1/\nu} \label{binder_prime_scal}
\eeq
thus giving an estimate of the $\nu$ critical index.

\section{Numerical Results}\label{results}

The first order transition, which is already quite weak at $h = 0$, gets weaker for negative $h$ 
values, so that we do not need to use algorithms specifically designed for strong first orders, like the multicanonical one.
While approaching the tricritical point autocorrelation times grow up, however, since we will perform our simulations mainly
in the first order region, this slowing down is not expected to be too significant for our study. Numerical 
simulations have thus been performed using a standard Metropolis algorithm.

Collected statistics have been of the order of $10^7\div10^8$ independent configurations
for all volumes and parameter sets explored; numerical simulations have been performed on GRID resources provided
by INFN.

\subsection{Discerning the critical behaviour from finite size scaling.}
\label{resultsA}

One way to discern between a first order and a second order critical behaviour\footnote{For the sake of simplicity we will speak 
of ``critical behaviour'' also for the case of first order transitions, although this is not completely appropriate.}
is to look at the distribution of physical observables, like the energy, 
at the transition point: that is expected to develop a double peak structure, 
in the thermodynamical limit, for a first order transition, while it stays single peaked in 
the second order case. In Figs.~\ref{histo0.0025} and \ref{histo0.01} we show two examples,
for $h = -0.0025$ and $h = -0.01$ respectively, where the situation is quite clear: 
$h = -0.0025$ clearly belongs to the first order region, while 
$h = -0.01$ appears to be on the second order side.

\begin{figure}[h!]
\includegraphics*[width=0.48\textwidth]{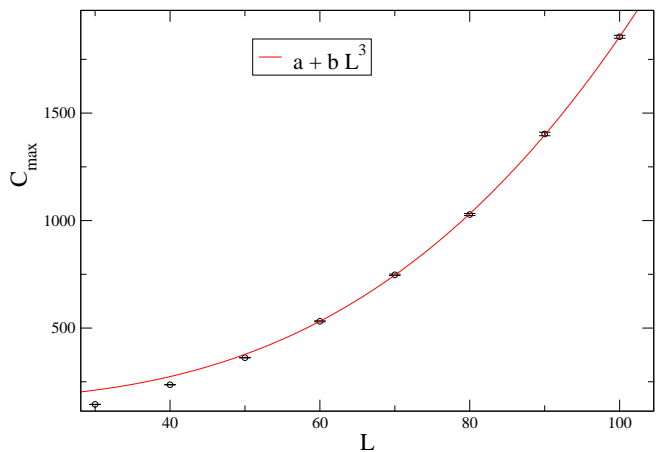}
\caption{Scaling of the specific heat peak with $L$ for $h = -0.0025$. $\chi^2/{\rm d.o.f.} \simeq 0.4$ 
(range of fit: $L>50$).}
\label{specheat00025}
\end{figure}

\begin{figure}[h!]
\includegraphics*[width=0.48\textwidth]{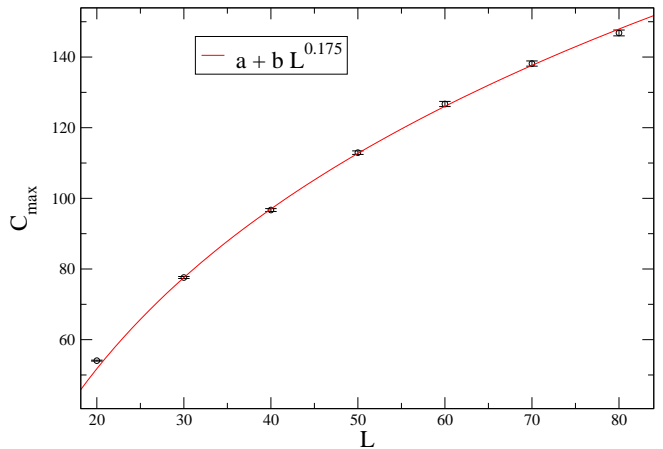}
\caption{Scaling of the specific heat peak with $L$ for $h = -0.01$. In this 
case the correct scaling with 3D Ising critical indexes is visible already
from moderate size lattices. $\chi^2/{\rm d.o.f.} \simeq 0.95 $ (range of fit: $L>20$)} 
\label{specheat0010}
\end{figure}

\begin{figure}[h!]
\includegraphics*[width=0.48\textwidth]{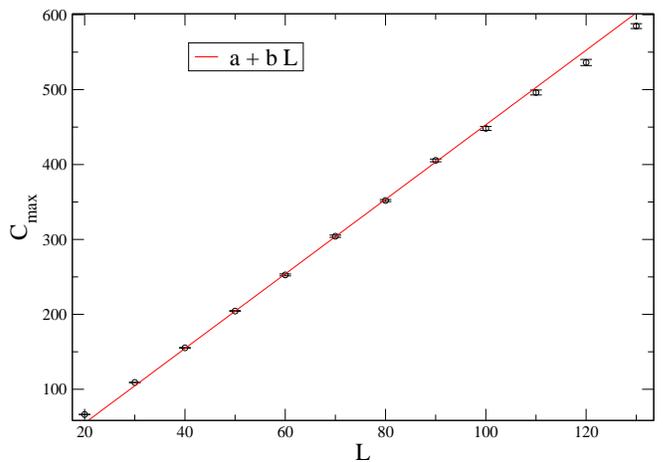}
\caption{Scaling of the specific heat peak with $L$ for $h = -0.005$. According
to our determination of $h_{\rm tric}$, the system belongs to the 3D Ising universality
class, however data scale according to tricritical indexes, with small 
deviations appearing only on the largest available volumes. $\chi^2/{\rm d.o.f.} \simeq 1.4$ (lattices with $L>100$
are not included in the fit).}
\label{specheat0005}
\end{figure}

Such conclusions are confirmed by looking at the scaling of the height of the specific heat peak. For 
$h = -0.025$ (see Fig.~\ref{specheat00025}) a cubic term in $L$, which is characteristic of first order, 
nicely fits the behaviour on the larger volumes.  For $h = -0.01$ (see Fig.~\ref{specheat0010}) the situation 
is also quite clear and data correctly scale according to 3D Ising critical indexes. 

What is less clear is the critical behaviour of the specific heat peak for $h = -0.005$, which 
is shown in Fig.~\ref{specheat0005}. Data scale linearly with $L$, \ie according to tricritical
indexes, for a large range of lattice sizes, with small deviations, going in the direction
of a smallest value of $\alpha/\nu$ (hence in the direction of the 3D Ising class), visible only on 
the largest sizes explored, $L > 100$. Our subsequent analysis will clearly show that for this value
of $h$ the system  belongs to the 3D Ising universality class, however it would have been difficult
to state that clearly from the scaling of the specific heat alone: tricritical indexes regulate
the system behavior till $L \sim 100$, completely masking the correct thermodynamical limit,
which would be evident only on much larger lattices. We expect the situation to be worse
and worse as one gets closer to $h_{\rm tric}$ (see the previous discussion in Sec.~\ref{setup}).

\begin{figure}[h!]
\includegraphics*[width=0.48\textwidth]{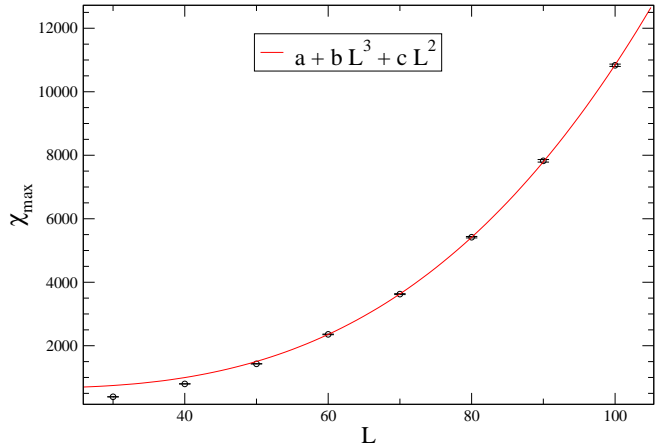}
\caption{Scaling with $L$ of the peak of the order parameter susceptibility for $h = -0.0025$. The contribution 
of two different terms, corresponding respectively to first order scaling and tricritical scaling, is needed to 
correctly fit our data. $\chi^2/{\rm d.o.f.} \simeq 0.45$ (range of fit: $L > 50$).}
\label{z2susc00025}
\end{figure}

\begin{figure}[h!]
\includegraphics*[width=0.48\textwidth]{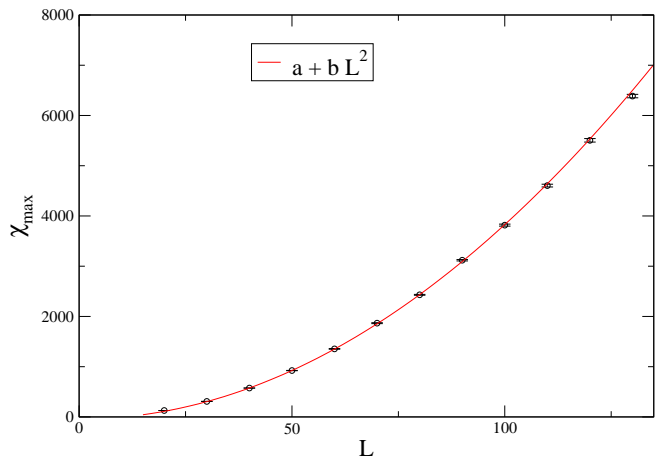}
\caption{Scaling with $L$ of the peak of the order parameter susceptibility 
for $h = -0.005$. In this case it is hard to disentangle tricritical 
from 3D Ising scaling, since $\gamma/\nu = 2$ in the first case and
$\gamma/\nu \simeq 1.968$ in the second case.  $\chi^2/{\rm d.o.f.} \simeq 1.4$ (range of fit: $L < 130$).} 
\label{z2susc0005}
\end{figure}

\begin{figure}[h!]
\includegraphics*[width=0.48\textwidth]{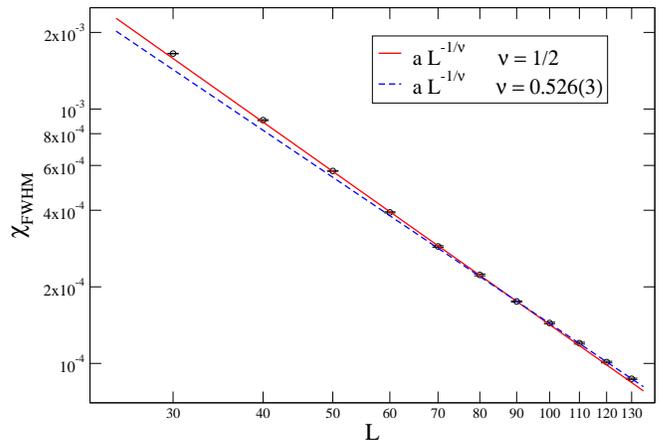}
\caption{Scaling with $L$ of the half height width of the peak of the order parameter susceptibility 
for $h = -0.005$. While till moderate size the scaling is compatible with a tricritical one, 
on larger volumes a deviation from the tricritical behaviour is clearly seen. $\chi^2/{\rm d.o.f.} \simeq 1.3$ 
for the small volumes ($L < 100$), $\chi^2/{\rm d.o.f.} \simeq 0.3$ for the larger ones ($L > 80$).}  
\label{z2suscFWHM0005}
\end{figure}

The situation is even more difficult when studying the scaling of the peak of the order parameter 
susceptibility, $\chi$, since in this case the relevant critical index, $\gamma/\nu$, practically 
coincides for the tricritical ($\gamma/\nu = 2$) and 3D Ising cases ($\gamma/\nu \simeq 1.963$). 

For $h = -0.0025$ one clearly sees a first order contribution (see Fig.~\ref{z2susc00025}). Notice however that, 
in order to correctly fit data, it is necessary to take into account also a small but non-zero contribution
proportional to $L^2$; this is the dominant term in the case of tricritical scaling, therefore 
we can interpret that as evidence for a non-negligible influence from a possibly close tricritical point.
The fit with the functional form $a+bL^3+cL^2$ gives the estimates for the parameters $a=710(70)$, $b=0.0139(3)$ and
$c=-0.37(3)$, meaning that the tricritical corrections to first order is about $20\%$ on the largest lattices explored
for this $h$ value.
For $h = -0.005$ there is no hope to discern between mean field and 3D Ising (see Fig.~\ref{z2susc0005}). 

A better probe in this case is furnished by the width of the susceptibility peak at half height, which is 
expected to scale like $L^{-1/\nu}$, since $1/\nu$ changes appreciably from mean field tricritical to 3D Ising 
(see Table~\ref{CRITEXP}). Data for the width are shown in Fig.~\ref{z2suscFWHM0005}: two different 
regimes are visible, the first for $L < 100$ regulated by the tricritical exponent ($\nu = 1/2$) and the second,
for larger lattices, where this index is sensibly larger, $\nu=0.526(3)$. We explicitly note that this value 
for $\nu$ is to be regarded just as an ``effective'', size dependent,  index interpolating between the 
tricritical ($\nu=1/2$) and Ising one ($\nu=0.63$).
That confirms what already found by looking at the scaling of the specific heat.

\begin{figure}[h!]
\includegraphics*[width=0.46\textwidth]{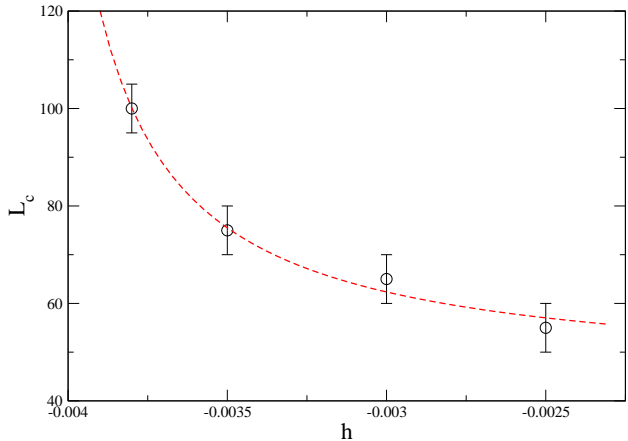}
\caption{On the vertical axis it is plotted the size $L_c$ such that for $L>L_c$ the scaling of the maxima of the energy
susceptibility is well described by a first-order scaling. The red line is a fit of the form $a+b/|h-h_{\rm tric}|$ 
where $h_{\rm tric}=-0.00415(3)$ (see Section \ref{tric_pos_section}).} 
\label{triscal_data}
\end{figure}

Let us summarize and comment the results contained in this subsection. Discerning the correct 
critical behaviour from the finite size scaling analysis of susceptibilities or other quantities 
may be a difficult task since, as expected, tricritical behavior masks the correct asymptotic scaling behaviour 
for some range of lattice sizes, which increases as we get closer to the tricritical field $h_{\rm tric}$
according to the tricritical crossover exponents, as summarized in Eq.~(\ref{critsize}). We have tried to 
verify quantitatively the prediction reported in Eq.~(\ref{critsize}) by estimating, for each value 
of $h$ on the first order side, the critical size $L_c$ such that for $L > L_c$ the scaling of the maxima
of the specific heat is well described by a first order scaling. Results are reported in 
Fig.~\ref{triscal_data}: Eq.~(\ref{critsize}) is well verified by 
using the value of $h_{\rm tric}$ obtained and reported 
in Section~\ref{tric_pos_section}.

The difficulties are generally larger on the second order side than on the first order one,
and we can easily understand why: the growth of susceptibilities is larger for 
mean field tricritical indexes than for 3D Ising critical indexes, hence a fake tricritical behavior 
can mask 3D Ising indexes for a large range of lattice sizes; on the other hand a first
order behaviour, which implies a faster growth of susceptibilities with respect to the 
tricritical one, is in general more easily detectable as a correction to tricritical behaviour 
starting from smaller lattice sizes.
As a last comment, we note that such difficulties make it preferable to look at the scaling of the specific
heat rather than at that of the order parameter, since the critical index regulating the growth
of the specific heat with $L$, $\alpha/\nu$, changes more drastically when going from first
order ($\alpha/\nu = 3$) to tricritical ($\alpha/\nu = 2$) and to 3D Ising ($\alpha\nu \simeq 0.175$).

\subsection{The latent heat, the order parameter gap and determination of $h_{\rm tric}$}
\label{tric_pos_section}

As explained in Section~\ref{setup}, we will now determine the parameters which fix
the strength of the first order transition taking place for small values of $|h|$, in order
to extrapolate the critical value $h_{\rm tric}$ at which the first order disappears. The parameters
are the latent heat, or equivalently the minimum of the Challa-Landau-Binder cumulant
defined in Eq.~(\ref{Bmin}), and the gap of the order parameter, which can be extracted
from the large volume limit of the maximum of its susceptibility $\chi$, see Eq.~(\ref{suscmax}).

\begin{figure}[h!]
\includegraphics*[width=0.47\textwidth]{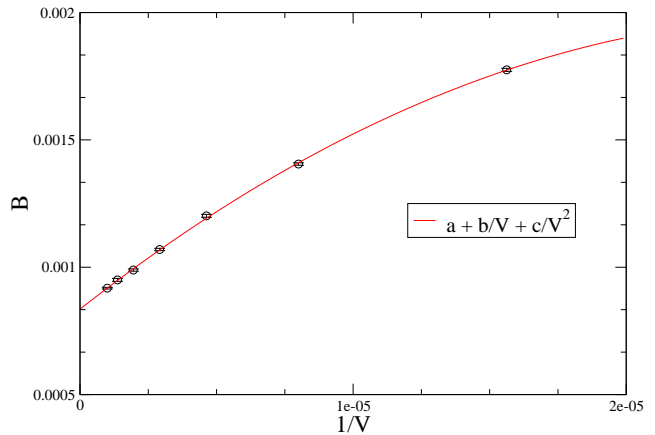}
\caption{Binder-Challa-Landau cumulant of the energy 
for $h = -0.0025 > h_{\rm tric}$. $B$ extrapolates to a non zero value
as $V \to \infty$. $\chi^2/{\rm d.o.f.} \simeq 1.4$.}
\label{Binder_0.0025}
\end{figure}

\begin{figure}[h!]
\includegraphics*[width=0.47\textwidth]{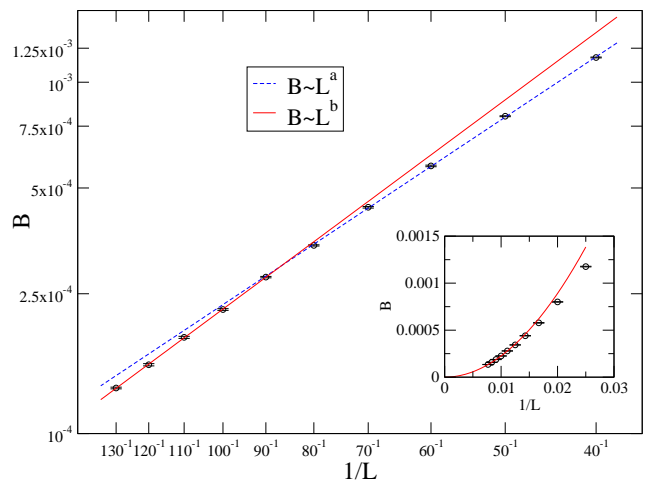}
\caption{As in Fig.~\ref{Binder_0.0025},  for $h = - 0.005 < h_{\rm tric}$. $B$ goes to zero but
the power law changes beyond a given size separating tricritical from 3D Ising scaling: $a=-1.77(1)$ and $b=-1.98(1)$.
$\chi^2/{\rm d.o.f.} \simeq 1.4$ ($L < 100$) and $\chi^2/{\rm d.o.f.} \simeq 0.4$ ($L > 80$) respectively.}
\label{Binder_0.005}
\end{figure}

In Fig.~\ref{Binder_0.0025} we show the quantity $B$ (see Eq.~(\ref{Bmin}) and the related discussion) as a function of $1/V$
for $h = -0.0025$. It clearly extrapolates to a non-zero value for $V \to \infty$, $a=8.35(4)\times 10^{-4}$, with
both $1/V$ and $1/V^2$ corrections visible in the range of explored volumes. For the same 
value of $h$ and using the same fit shown in Fig.~\ref{z2susc00025}, from the coefficient
of the cubic term in $L$ we extract the order parameter gap, $\Delta^2 = 1.689(5)\times 10^{-3}$. 
In Fig.~\ref{Binder_0.005} instead we show the case $h = -0.005$, together with a power law fit $B\propto L^{a}$. If we try instead
 $B=b_0+L^a$ we get for $b_0$ the result consistent with zero shown in Tab.~\ref{table_gap}, indicating that no latent heat is present.

\begin{table}
\begin{tabular}{|l|l|l|}
\hline 
\rule{0mm}{3.4mm}\hfill $h$ \hfill{}     & \hfill $B$ \hfill{} & \hfill $\Delta^2$ \hfill{} \\ \hline 
\rule{0mm}{3.4mm}  $-0.002$   & $1.17(2)\times 10^{-3}$  &  $1.45(2)\times 10^{-2}$   \\ \hline
\rule{0mm}{3.4mm}  $-0.0025$  & $8.29(6)\times 10^{-4}$  &  $1.28(7)\times 10^{-2}$   \\ \hline
\rule{0mm}{3.4mm}  $-0.003$   & $5.1(1) \times 10^{-4}$  &  $7.8(3) \times 10^{-3}$   \\ \hline
\rule{0mm}{3.4mm}  $-0.0035$  & $2.78(1)\times 10^{-4}$  &  $5.6(3) \times 10^{-3}$   \\ \hline
\rule{0mm}{3.4mm}  $-0.0038$  & $1.7(2) \times 10^{-4}$  &  $2.7(4) \times 10^{-3}$   \\ \hline
\rule{0mm}{3.4mm}  $-0.005$   & $5(11) \times 10^{-6}$   &  $-2(2.8) \times 10^{-4}$   \\ \hline
\end{tabular}
\caption{Estimated values for the thermodynamical limit of \(B\) and \(\Delta^2\).}\label{table_gap}
\end{table}

\begin{figure}[h!]
\includegraphics*[width=0.5\textwidth]{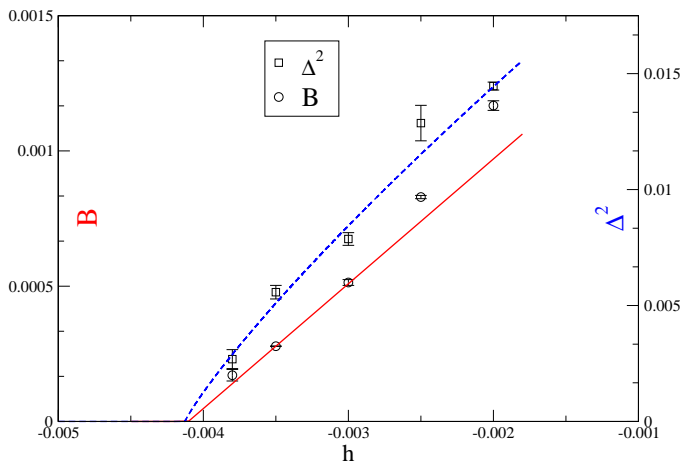}
\caption{Extrapolation of the Binder cumulant and of the order parameter gap
in order to extract the tricritical value of the magnetic field $h_{\rm tric}$.}
\label{endpoint}
\end{figure}

We have applied the same procedure to all values of $h$ where the first order transition
is clearly detectable on the explored volumes, obtaining the values for $B$
and the gap reported in Table~\ref{table_gap}. From those values, and using the expected
behaviors reported in Eqs.~(\ref{deltae_beh}) and (\ref{delta_beh}), we can fit the value
of $h_{\rm tric}$ from both quantities. Results are reported in Fig.~\ref{endpoint}: we obtain
$h_{\rm tric} = -0.00410(5)$ from the extrapolated minimum of the cumulant, and 
$h_{\rm tric} = -0.00412(7)$ from the order parameter gap. The two values are in perfect
agreement with each other and with the outcome of the finite size scaling analysis reported above;
however the finite size scaling analysis alone would have not been able to locate 
$h_{\rm tric}$ with such precision. 
We also notice that our determination for $h_{\rm tric}$ is in good agreement with the results reported 
in Refs.~\cite{pottsim2} and \cite{rwep2} (Fig.~5 in both references), whose estimate was\footnote{P.~de Forcrand, private communication.}
$h_{\rm tric}=-0.00445(20)$. 

As an alternative, independent way to locate $h_{\rm tric}$, we have studied the cumulant $U_4$ defined in
Eq.~(\ref{z2_binder}). The theoretical expectation is that increasing the lattice size $U_4\to 1$ for 
$h>h_{\rm tric}$, $U_4\to U_4^{\rm tric}$ for $h=h_{\rm tric}$ and $U_4\to U_4^{\rm Ising}$ for $h<h_{\rm tric}$. 
In particular the cumulants calculated on different lattices are expected to intersect at the tricritical point,
with a  slope increasing as $L^{1/\nu}$ (see Eq. \ref{binder_prime_scal}), with $\nu=1/2$.

\begin{figure}[h!]
\includegraphics*[width=0.49\textwidth]{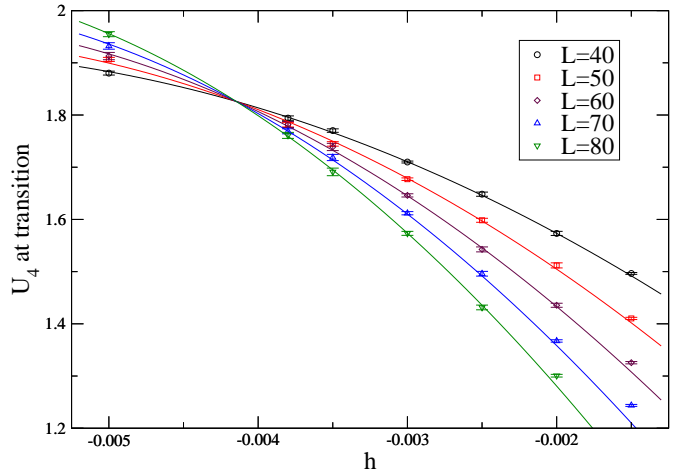}
\caption{Binder cumulant of the order parameter at the critical temperature, 
as defined in Eq.~\ref{z2_binder}. The curves at different volumes
intersect at $h_{\rm tric}$.}
\label{z2binder}
\end{figure}

\begin{figure}[h!]
\includegraphics*[width=0.49\textwidth]{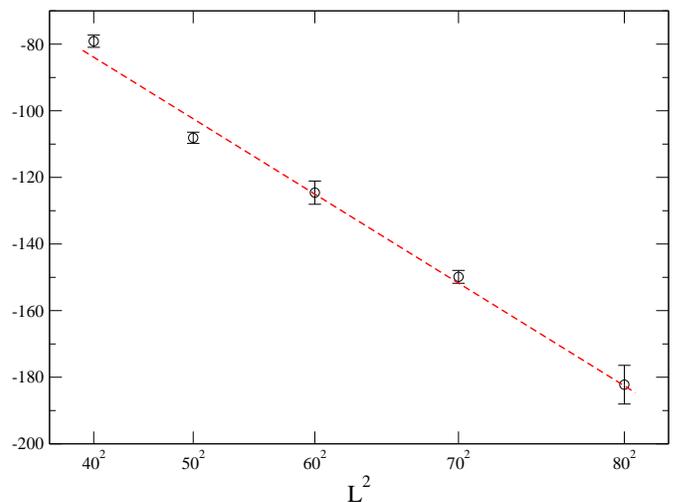}
\caption{Slope of the Binder cumulant of the order parameter at the critical temperature. The line is a linear fit.}
\label{z2binderprime}
\end{figure}

Numerical results for $U_4$ are reported in Fig.~\ref{z2binder} and Table \ref{table_z2binder}; the location of the 
intersection point is determined by using the method exposed in \cite{crossing}, \S III.B: a scaling law of the form
$U_4=f((h-h_{\rm tric})L^y)$ is assumed and, since we are sufficiently close to the tricritical point, we can develop
$f(x)$ in power series around $x=0$ (also scaling corrections are usually to be taken into account, see the discussion
in \cite{crossing}); a fit is then performed in the expansion parameters taking into account the data measured at different 
$L$ values. By using data for $-0.005\le h < -0.002$ and $L\ge 40$ the fit has 19 d.o.f. and $\chi^2/{\rm d.o.f.}\simeq
0.8$ The estimated location of the tricritical point is $h_{\rm tric}=-0.00415(3)$. In Fig.~\ref{z2binderprime} it is 
shown that the derivative of the cumulant scales with the expected critical index.

\begin{table}
\begin{tabular}{|l|l|l|l|l|l|}
\hline 
\hfill $h$ \hfill{} & \hfill $L=40$ \hfill{} &  \hfill $L=50$ \hfill{} &  \hfill $L=60$ \hfill{} &  
\hfill $L=70$ \hfill{} & \hfill $L=80$ \hfill{}  \\ \hline 
$-0.0015$  &  $1.496(2)$ &   $1.410(2)$ &   $1.325(2)$ &   $1.244(2)$ &   \hfill $\cdots$ \hfill {} \\ \hline
$-0.002$   &  $1.573(4)$ &   $1.512(5)$ &   $1.435(4)$ &   $1.367(2)$ &   $1.301(3)$ \\ \hline
$-0.0025$  &  $1.648(4)$ &   $1.598(4)$ &   $1.542(5)$ &   $1.496(4)$ &   $1.431(5)$ \\ \hline
$-0.003$   &  $1.710(2)$ &   $1.677(2)$ &   $1.646(3)$ &   $1.612(3)$ &   $1.573(4)$ \\ \hline
$-0.0035$  &  $1.770(3)$ &   $1.745(4)$ &   $1.739(7)$ &   $1.719(6)$ &   $1.691(7)$ \\ \hline
$-0.0038$  &  $1.794(4)$ &   $1.781(5)$ &   $1.782(6)$ &   $1.769(5)$ &   $1.761(6)$ \\ \hline
$-0.005$   &  $1.880(4)$ &   $1.906(4)$ &   $1.913(7)$ &   $1.932(6)$ &   $1.955(5)$ \\ \hline
$-0.01$    &  $2.080(6)$ &   $2.071(6)$ &   $2.121(7)$ &   $2.116(8)$ &   $2.123(9)$ \\ \hline
\end{tabular}
\caption{\(U_4\) values at the transition for different lattice sizes and magnetic field.}\label{table_z2binder}
\end{table}

\subsection{Critical temperature}

We conclude the presentation of our numerical results by analyzing the behaviour
of the critical temperature as a function of $h$, $\beta_c(h)$. Our determinations of 
$\beta_c$ are summarized in Table~\ref{betac} and they have been obtained by using the
number of phase criterion for first order transitions (\cite{Janke}), while for the 
second order ones the crossing point of the order parameter cumulant was used 
(see \eg \cite{Binder2}).

\begin{table}[!t]
\begin{tabular}{|l|l|}
\hline 
\rule{0mm}{3.4mm}\hfill $h$ \hfill{}  & \hfill $\beta_c$ \hfill{} \\ \hline 
\rule{0mm}{3.4mm}  $-0.0015$  & $0.549537(4)$   \\ \hline
\rule{0mm}{3.4mm}  $-0.002$   & $0.549237(2)$   \\ \hline
\rule{0mm}{3.4mm}  $-0.0025$  & $0.548942(1)$   \\ \hline
\rule{0mm}{3.4mm}  $-0.003$   & $0.548652(1)$   \\ \hline
\rule{0mm}{3.4mm}  $-0.0035$  & $0.548358(3)$   \\ \hline
\rule{0mm}{3.4mm}  $-0.0038$  & $0.548199(2)$   \\ \hline
\rule{0mm}{3.4mm}  $-0.005$   & $0.5475152(6)$  \\ \hline
\rule{0mm}{3.4mm}  $-0.01$    & $0.545071(8)$   \\ \hline
\rule{0mm}{3.4mm}  $-0.5$     & $0.484166(5)$   \\ \hline
\rule{0mm}{3.4mm}  $-1.0$     & $0.465188(4)$   \\ \hline
\rule{0mm}{3.4mm}  $-1.5$     & $0.45576(1)$    \\ \hline
\rule{0mm}{3.4mm}  $-2.0$     & $0.450591(4)$   \\ \hline
\end{tabular}
\caption{Estimated values for $\beta_c$ at fixed $h$.}\label{betac}
\end{table}

\begin{figure}[h!]
\includegraphics*[width=0.49\textwidth]{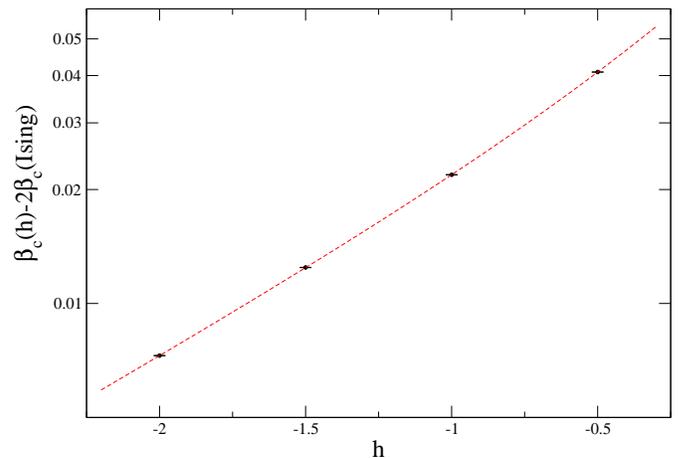}
\caption{Plot of $\beta_c(h)-2\beta_c({\rm Ising})$. The line is the result of a fit with the function in 
Eq.~(\ref{high_beta_fit}).}
\label{betac_large}
\end{figure}

We expect that for large negative values of $h$ the state coupled to the magnetic
field disappears from the system dynamics, which then becomes completely equivalent to that 
of a 3D Ising system. That must be visible from the behaviour of $\beta_c(h)$ which should 
approach two times\footnote{This multiplicative factor is caused by a different normalization in the Hamiltonians: 
the Ising one is usually written as a sum of terms $\beta \sigma_i \sigma_j=\beta (2\delta_{\sigma_i \sigma_j}-1)$.} 
$\beta_c({\rm Ising}) = 0.2216546(10)$ (\cite{betaising}) as $h \to -\infty$. In Fig.~\ref{betac_large} we 
show the quantity $\beta_c(h) - 2\, \beta_c({\rm Ising})$, in the regime of large $|h|$, which is expected to vanish  
in the same limit; indeed we have verified that the functional behaviour expected from a strong coupling expansion
\beq
\beta_c(h) - 2\beta_c({\rm Ising}) = b_1 e^h + b_2 e^{2h} + b_3 e^{3h}  \label{high_beta_fit}
\eeq
fits our data with $b_1 = 0.0514(1)$, $b_2 = 0.0185(5)$, $b_3 = 0.0150(5) $ and  
$\tilde\chi^2/{\rm d.o.f.} = 2.8$. The agreement is reasonable taking into account that our data are very accurate and 
we truncate the strong coupling series just to third order. 

\begin{figure}[h!]
\includegraphics*[width=0.49\textwidth]{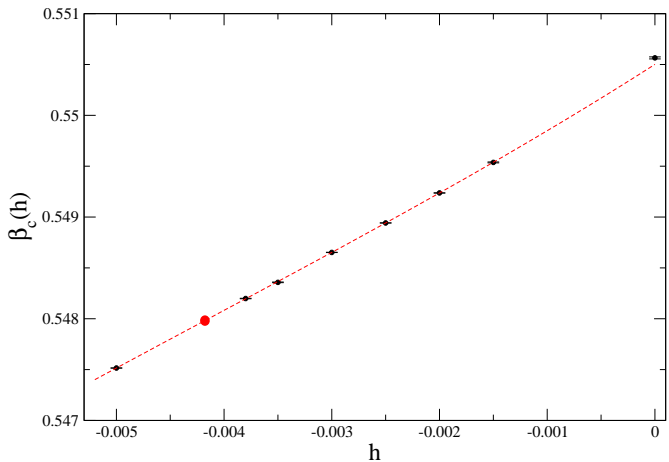}
\caption{Plot of $\beta_c(h)$. The line is the result of a fit with the function in Eq.~\ref{low_beta_fit}. The point at
$h=0$ is the result of \cite{Janke} and is not included in the fit. The (red in color) spot on the fitted line corresponds
to our location of the tricritical point.}
\label{betac_small}
\end{figure}

In the opposite limit of small values of $h$, as shown in Fig.~\ref{betac_small}, 
we have been able to fit the 
$\beta(h)$ dependence by a third order polynomial in $h$
\beq
\beta_c(h) = \tilde{b}_0 + \tilde{b}_1 h + \tilde{b}_2h^2 + \tilde{b}_3h^3 \label{low_beta_fit}
\eeq
with $\tilde{b}_0=0.550500(45)$ $\tilde{b}_1 = 0.676(44)$, $\tilde{b}_2 = 26(13)$, $\tilde{b}_3 = 2100(1400) $ and  
$\tilde\chi^2/{\rm d.o.f.} = 2.5$. We notice that $b_0$ gives an estimate of the critical point position of the Potts 
model without external field compatible with the known result $\beta_c(h=0)=0.550565(10)$ obtained in \cite{Janke} 
and that the slope of $\beta_c (h)$ at $h = 0^-$ is different from the one observed on the positive $h$ side \cite{Karsch}.

Finally, fitting data for $\beta_c(h)$ around $h_{\rm tric}$, 
we can estimate also the temperature location of the tricritical point and state 
$(\beta_{\rm tric},h_{\rm tric}) = (0.5480(1),-0.00415(3))$. 

\section{Conclusions}\label{concl}

We have investigated the critical properties of the three-dimensional three-state Potts
model as a function of a negative magnetic field coupled to one of the three spin states.
In this case the system possesses a residual exact symmetry, which gets spontaneously
broken at a critical coupling $\beta_c(h)$, which approaches twice the critical 
coupling of the 3D Ising model for $|h| \to \infty$.
In particular, we have determined the tricritical value $h_{\rm tric}$ at which 
the finite temperature first order transition, taking place for null or small values
of $|h|$, turns into a second order transition in the universality class of the 
3D Ising model.

We have shown that, in proximity of the tricritical field
$h_{\rm crit}$, it is difficult to determine the critical behavior from
the finite size scaling of susceptibilities alone, since, 
at a given distance from $h_{\rm crit}$, tricritical scaling 
masks the correct critical indexes up to a given lattice size $L_{\rm max}$, which is regulated 
by tricritical crossover exponents ($L_{\rm max} \propto |h - h_{\rm tric}|^{-1}$ in our case).
A better strategy is to determine parameters which fix the strength of the first
order region, like the order parameter gap $\Delta$ or the latent heat 
$\Delta_E$, and to determine $h_{\rm tric}$ as the value of $h$ at which these
parameters extrapolate to zero. We also showed that the order parameter cumulant 
is another very useful quantity to look at. In this way we have obtained the quite accurate
estimate $h_{\rm tric} = -0.00415(3)$, which is in agreement with previous
determinations reported in Refs.~\cite{pottsim2,rwep2}. Although in this work the cumulant method has proved to be the 
most efficient, which of the studied methods is to be preferred to locate a tricritical point 
should be model dependent.

Our results may be useful in lattice QCD studies aimed at
determining the order and universality class of the Roberge-Weiss endpoint,
and the values of the tricritical masses separating the second
order from the first order regions both for two~\cite{rwep, bcds} and
three~\cite{rwep2} degenerate flavors. In particular we expect that distinguishing
the correct critical behavior on feasible lattice sizes will be increasingly difficult
as we approach the tricritical masses, the specific heat being anyway more sensitive 
than the order parameter susceptibility. An accurate determination of the tricritical
masses may be based, for instance, on a careful study of the transition strength as a function of $h$
in the first order regions.

\section*{Acknowledgments}

We thank V.~Alba, G.~Cossu, P.~de~Forcrand, A.~Di Giacomo, F.~Sanfilippo and E.~Vicari for useful discussions. 
Numerical simulations have been performed on GRID
resources provided by INFN.


\begin{thebibliography}{9}

\bibitem{potts} R.~B.~Potts, 
  Proc. Cambridge Philos. Soc. {\bf 48}, 106 (1952).

\bibitem{wu} F.~Y.~Wu, 
  Rev. Mod. Phys. {\bf 54}, 235 (1982).

\bibitem{svya} B. Svetitsky and L.G. Yaffe,  
  Nucl. Phys. B {\bf 210}, 423 (1982).

\bibitem{gavai} R.~V.~Gavai, F.~Karsch, and B.~Petersson, 
  Nucl. Phys. B {\bf 322}, 738 (1989).

\bibitem{fuku} M.~Fukugita, H.~Mino, M.~Okawa, and A.~Ukawa,
  J. Stat. Phys. {\bf 59}, 1397 (1990).

\bibitem{alves} N.~A.~Alves, B.~A.~Berg and R.~Villanova, 
  Phys. Rev. B {\bf 43}, 5846 (1991).

\bibitem{schmidt} M.~Schmidt, 
  Z. Phys. B {\bf 95}, 327 (1994).

\bibitem{Janke} 
  W. Janke and R. Villanova, 
  Nucl. Phys. B {\bf 489}, 679 (1997) 
  [arXiv:hep-lat/9612008].

\bibitem{Karsch} F. Karsch and S. Stickan,
  Phys. Lett. B {\bf 488}, 319 (2000) 
  [arXiv:hep-lat/0007019].

\bibitem{caselle1}
  M.~Caselle, M.~Hasenbusch, P.~Provero and K.~Zarembo,
  Nucl.\ Phys.\  B {\bf 623}, 474 (2002)
  [arXiv:hep-th/0103130].

\bibitem{pottsim1}
  M.~G.~Alford, S.~Chandrasekharan, J.~Cox and U.~J.~Wiese,
  Nucl. Phys. B {\bf 602}, 61 (2001)
  [arXiv:hep-lat/0101012].

\bibitem{pottsim2}
  S.~Kim, Ph.~de Forcrand, S.~Kratochvila and T.~Takaishi,
  PoS {\bf LAT2005}, 166 (2006)
  [arXiv:hep-lat/0510069].

\bibitem{caselle2}
  M.~Caselle, G.~Delfino, P.~Grinza, O.~Jahn and N.~Magnoli,
  J. Stat. Mech.  {\bf 06}, P03008 (2006)
  [arXiv:hep-th/0511168].

\bibitem{papa} R.~Falcone, R.~Fiore, M.~Gravina, and A.~Papa, 
  Nucl. Phys. B {\bf 767}, 385 (2007)
  [arXiv:hep-lat/0612016]

\bibitem{babe} A.~Bazavov and B.~A.~Berg, 
  Phys. Rev. D {\bf 75}, 094506 (2007)
  [arXiv:hep-lat/0702018].

\bibitem{babedu}
  A.~Bazavov, B.~A.~Berg and S.~Dubey,
  Nucl.\ Phys.\  B {\bf 802}, 421 (2008)
  [arXiv:0804.1402 [hep-lat]].

\bibitem{degrand0}
  T.~DeGrand, R.~Hoffmann,
  JHEP {\bf 0702}, 022 (2007)
  [arXiv:hep-lat/0612012].

\bibitem{degrand}
  T.~DeGrand, R.~Hoffmann, J.~Najjar,
  JHEP {\bf 0801}, 032 (2008)
  [arXiv:0711.4290 [hep-lat]].

\bibitem{lucini}
  B.~Lucini, A.~Patella and C.~Pica,
  Phys. Rev. D {\bf 75}, 121701 (2007)
  [arXiv:hep-th/0702167].

\bibitem{lucini2}
  B.~Lucini, A.~Patella,
  Phys. Rev. D {\bf 79}, 125030 (2009)
  [arXiv:0904.3479 [hep-th]].


\bibitem{asv}
  A.~Armoni, M.~Shifman and G.~Veneziano,
  Phys. Rev. Lett.  {\bf 91}, 191601 (2003)
  [arXiv:hep-th/0307097].

\bibitem{uy2006}
  M.~Unsal and L.~G.~Yaffe,
  Phys. Rev. D {\bf 74}, 105019 (2006)
  [arXiv:hep-th/0608180].

\bibitem{rw}
  A.~Roberge, N.~Weiss,
  Nucl. Phys. B {\bf 275}, 734 (1986).

\bibitem{sqgp}
  M.~D'Elia, F.~Di Renzo and  M.P.~Lombardo,
  Phys. Rev. D {\bf 76}, 114509 (2007)
  [arXiv:0705.3814 [hep-lat]].

\bibitem{Kouno:2009bm}
  H.~Kouno, Y.~Sakai, K.~Kashiwa and M.~Yahiro,
  J. Phys. G {\bf 36}, 115010 (2009)
  [arXiv:0904.0925 [hep-ph]].

\bibitem{rwep}
  M.~D'Elia and F.~Sanfilippo,
  Phys. Rev. D {\bf 80}, 111501 (2009)
  [arXiv:0909.0254 [hep-lat]].

\bibitem{rwep2}
  P.~de Forcrand and O.~Philipsen,
  Phys. Rev. Lett. {\bf 105}, 152001 (2010)
  [arXiv:1004.3144 [hep-lat]].

\bibitem{aarts}
  G.~Aarts, S.~P.~Kumar and J.~Rafferty,
  JHEP {\bf 1007}, 056 (2010)
  [arXiv:1005.2947 [hep-th]].

\bibitem{ssky}
  Y.~Sakai, T.~Sasaki, H.~Kouno, M.~Yahiro
  Phys. Rev. D 82, 076003 (2010)
  [arXiv:1006.3648 [hep-ph]].

\bibitem{bcds}
  C.~Bonati, G.~Cossu, M.~D'Elia, F.~Sanfilippo
  [arXiv:1011.4515 [hep-lat]].

\bibitem{landau} 
  L.~D.~Landau and E.~M.~Lifshitz, ``Statistical Physics, Part 1'', Butterworth Heinemann (1980).

\bibitem{pelissvic}
  A.~Pelissetto and E.~Vicari, 
  Phys. Rep. {\bf 368}, 549 (2002)
  [arXiv:cond-mat/0012164].

\bibitem{BinderDeutsch}
  K.~Binder and H.~P.~Deutsch,
  Europhys. Lett. {\bf 18}, 667 (1992).

\bibitem{Cardy}
  J.~Cardy, ``Scaling and Renormalization in Statistical Physics'', Cambridge University Press (2003).

\bibitem{LawSarb}
  I.~D.~Lawrie and S.~Sarbach, \emph{Theory of Tricritical Points}, in C.~Domb, J.~L.~Lebowitz (eds.)
  ``Phase transitions and critical phenomena, vol. 11'', Academic Press (1987).

\bibitem{Challa}
  M.~S.~S.~Challa, D.~P.~Landau and K.~Binder,
  Phys. Rev.  B {\bf 34}, 1841 (1986).

\bibitem{LeeKosterlitz} J.~Lee, J.~M.~Kosterlitz,
  Phys. Rev. B {\bf 43}, 3265 (1991).

\bibitem{Sheehy}
  D.~E~Sheehy, 
  Phys. Rev. A {\bf 79}, 033606 (2009)
  [arXiv:0807.0922 [cond-mat]].

\bibitem{Binder1}
  K.~Binder, 
  Phys. Rev. Lett. {\bf 47}, 693 (1981).

\bibitem{Binder2}
  K.~Binder,
  Z. Phys. B - Condensed Matter {\bf 43}, 119 (1981).

\bibitem{betaising}
  H.~W.~J.~Bl\"{o}te, E.~Luijten and J.~R.~Heringa, 
  J. Phys. A: Math. Gen. {\bf 28}, 6289 (1995)
  [arXiv:cond-mat/9509016].

\bibitem{crossing}
  M.~Hasenbusch, F.~Parisen Toldin, A.~Pelissetto and E.~Vicari,
  Phys. Rev. E {\bf 77}, 051115 (2008) 
  [arXiv:0803.0444 [cond-mat.dis-nn]].


\end{thebibliography}
\end{document}